\documentclass[preprint,12pt]{elsarticle}

\usepackage{amssymb}
\usepackage{amsmath}
\usepackage{color,graphicx}
\usepackage{hyperref}

\journal{Physics Reports}

\begin{document}

\begin{frontmatter}

%% Title, authors and addresses

%% use the tnoteref command within \title for footnotes;
%% use the tnotetext command for theassociated footnote;
%% use the fnref command within \author or \affiliation for footnotes;
%% use the fntext command for theassociated footnote;
%% use the corref command within \author for corresponding author footnotes;
%% use the cortext command for theassociated footnote;
%% use the ead command for the email address,
%% and the form \ead[url] for the home page:
%% \title{Title\tnoteref{label1}}
%% \tnotetext[label1]{}
%% \author{Name\corref{cor1}\fnref{label2}}
%% \ead{email address}
%% \ead[url]{home page}
%% \fntext[label2]{}
%% \cortext[cor1]{}
%% \affiliation{organization={},
%%             addressline={},
%%             city={},
%%             postcode={},
%%             state={},
%%             country={}}
%% \fntext[label3]{}

\title{Parton distribution functions and strong coupling constant: MSHT} %% Article title

%% use optional labels to link authors explicitly to addresses:
%% \author[label1,label2]{}
%% \affiliation[label1]{organization={},
%%             addressline={},
%%             city={},
%%             postcode={},
%%             state={},
%%             country={}}
%%
%% \affiliation[label2]{organization={},
%%             addressline={},
%%             city={},
%%             postcode={},
%%             state={},
%%             country={}}

\author[affi1,affi2]{T. Cridge}\author[affi3]{L. A. Harland-Lang}\author[affi3]{M. Reader}\author[affi3]{R. S. Thorne}

%% Author affiliation
\affiliation[affi1]{organization={Elementary Particle Physics, University of Antwerp},%Department and Organization
            addressline={Groenenborgerlaan 171}, 
            city={2020 Antwerp},
%            postcode={15771}, 
%%          state={},
            country={Belgium}}

\affiliation[affi2]{organization={Department of Physics and Astronomy, University of Manchester},
            addressline={Oxford Rd}, 
            city={Manchester},
            postcode={M13 9PL}, 
%%          state={},
            country={United Kingdom}}

\affiliation[affi3]{organization={Department of Physics and Astronomy, University College London},
            addressline={Gower St}, 
            city={London},
            postcode={WC1E 6BT}, 
%%          state={},
            country={United Kingdom}}

%% Abstract
\begin{abstract}
We present an overview of the MSHT family of global PDF fits. We describe the underlying methodology, including parameterisation choice, theoretical inputs and datasets included in the fit. We describe the impact of recent theoretical advances in the inclusion of higher order QCD and EW corrections on the fit and their implications for phenomenology. We discuss recent methodological improvements and ongoing studies relating to closure tests of the parameterisation flexibility and the inclusion of errors on errors for experimental data. We in addition discuss recent results on the determination of the strong coupling from the MSHT fits.
\end{abstract}

%%Graphical abstract
%\begin{graphicalabstract}
%\includegraphics{grabs}
%\end{graphicalabstract}

%%Research highlights
%\begin{highlights}
%\item Research highlight 1
%\item Research highlight 2
%\end{highlights}

%% Keywords
\begin{keyword}
QCD, Electroweak, Parton Distribution Functions
%% keywords here, in the form: keyword \sep keyword

%% PACS codes here, in the form: \PACS code \sep code

%% MSC codes here, in the form: \MSC code \sep code
%% or \MSC[2008] code \sep code (2000 is the default)

\end{keyword}

\end{frontmatter}

%% Add \usepackage{lineno} before \begin{document} and uncomment 
%% following line to enable line numbers
%% \linenumbers

%% main text
%%

%% Use \section commands to start a section
%\section{MSHT PDFs}
%\label{sec1}
%% Labels are used to cross-reference an item using \ref command.

%Section text. See Subsection \ref{subsec1}.

%% Use \section commands to start a subsection.
%\section{MSHT PDFs}
%\label{subsec1}

\section{History of the PDFs}

The latest sets of parton distribution functions (PDFs) from the MSHT group are the MSHT20 \cite{Bailey:2020ooq} sets, which are presented at 
LO, NLO and NNLO, and most recently also at approximate N$^3$LO (aN$^3$LO)
\cite{McGowan:2022nag}. 
These supersede the MMHT14 PDFs \cite{Harland-Lang:2014zoa}, 
which were the successor to MSTW08~\cite{Martin:2009iq}, which in turn derived from the earlier MRST and MRS studies~\cite{Martin:1987vw,Martin:1994kn,Martin:1998sq}. The MRS analysis in \cite{Martin:1994kn} was the first truly global analysis, fitting to fixed target, DIS and hadroproduction data to constrain the 
PDFs as precisely as possible. Subsequent releases have all built on this approach, but with significant advances due to improvements in both theory and experiment, as well as methodology.
The MRST98 release~\cite{Martin:1998sq} was the first PDF set to include a full treatment of heavy flavours within the GM--VFNS developed in~\cite{Thorne:1997ga}, and the MRST02 release~\cite{Martin:2002aw} included a full treatment of PDF errors. 
The MRST04~\cite{Martin:2004ir} set went to NNLO for the first time, although earlier fits based on approximate NNLO treatments gave similar results and a proper treatment of the PDF discontinuities which occur at NNLO across heavy flavour thresholds was presented later in~\cite{Martin:2007bv}. MSTW08~\cite{Martin:2009iq} included a global fit to a variety of DIS data from HERA and fixed proton and nuclear targets, fixed target Drell--Yan and dimuon production and $W$, $Z$ and jet production at the Tevatron. Fits were performed at up to NNLO with an up--to--date heavy flavour scheme applied.
This provided a PDF set for use at the LHC, which
began operation soon after the release.
These PDFs were updated in the MMHT14~\cite{Harland-Lang:2014zoa} set, which included for the first time LHC data on $W$, $Z$, $t\overline{t}$ and jet production, as well as updated HERA and Tevatron data. 

\section{The MSHT20 PDFs.}

The acronym MSHT used for our most recent PDF sets is now a permanent naming convention,
standing for Mass Scheme Hessian Tolerance, and it incorporates some of the
central and enduring features of our approach to PDF fitting, which we will expand on below. 
The MSHT20 analysis includes new theoretical developments, and an extended parameterisation compared to MMHT14, in particular for $\bar d/\bar u$ and the strange quark. Consequently it contains more eigenvector sets, offering a more detailed evaluation of the PDF uncertainty. 
There is much new data, largely from the LHC, but also final HERA and Tevatron data sets. Almost all cross sections are included at NNLO in QCD perturbation theory. 
The fit quality is generally very good, but there are some places where it is difficult to obtain a good fit quality, and there are tensions between some data sets. It is found that NLO QCD is clearly no longer sufficient to provide an acceptable fit quality and hence to provide 
precise predictions. For the LO PDF set the fit quality is very poor.  

As in the MMHT14~\cite{Harland-Lang:2014zoa} analysis, the heavy flavour component of DIS is obtained using 
a general mass variable flavour scheme based on 
the TR scheme~\cite{Thorne:1997ga,Thorne:2006qt}, and in particular, the ``optimal'' choice~\cite{Thorne:2012az} which improves the smoothness of the transition region. Deuteron and heavy nuclear corrections are applied where appropriate, the former accounting for the binding of the proton and neutron within the deuteron. Deuteron corrections are modeled using $4$ unconstrained parameters which are determined, including uncertainties, from the PDF fit. There is very good agreement with the form used in the CJ12~\cite{Owens:2012bv} analysis, in particular the CJ12mid prediction.
The heavy nuclear corrections use the same parameterisations~\cite{deFlorian:2011fp} as MMHT14 with 
the fit allowing an additional penalty-free freedom of order $1\%$. 

With the exception of some of the oldest fixed-target DIS data sets, the  fits are performed using systematic uncertainties if possible, using either nuisance parameters (the preferred method) or the correlation matrix provided, and statistical correlations are also applied whenever available. In general, if both are available, we use the absolute cross sections in the fit in preference to normalised data in order to avoid loss of information. For the DIS data a cut of 2 ${\rm GeV}^2$ in $Q^2$ and a cut of 15 ${\rm GeV}^2$ in $W^2$ are imposed to avoid regions with sensitivity to higher twist corrections. The impact of these cuts and corrections is the topic of a very recent study~\cite{Harland-Lang:2025wvm}. 

The analysis includes many recently completed NNLO calculations.
We use the NNLO calculations for dimuon 
production~\cite{Berger:2016inr} in charged current DIS, where the hard cross section correction is negative at lower $x$, resulting in an increase of the strange quark at NNLO in order to fit to these data. This relieves some of the tension between the dimuon 
data~\cite{NuTeV:2001dfo} and LHC $W,Z$ data~\cite{ATLAS:2016nqi, ATLAS:2017rue,ATLAS:2019fgb}, the latter preferring a larger strange quark. 
The $D\to \mu$ branching ratio, which is required in the fit to dimuon production, is determined using the measurement of~\cite{Bolton:1997pq}, which is not derived from dimuon production data; the experimental uncertainty on this is propagated in the fit, providing a large contribution to the strangeness uncertainty when fitting to these data. Nearly all other data use theoretical 
calculations at full NNLO precision, in particular the \texttt{NNLOJET} cross-section calculations~\cite{Currie:2016bfm} are used for all LHC jet data, i.e.  
inclusive jet production at 2.76, 7 and 8~TeV, using the larger available 
jet radius, e.g. $R=0.6,0.7$. Older Tevatron jet  data (which carry little weight in the fit)
are still included,  but using the threshold approximation for NNLO~\cite{Kidonakis:2000gi} - which is a good approximation 
for these data. 
CMS $7$~TeV$~W+c$ data~\cite{CMS:2013wql} had only the NLO theory available for the specific 
measurement, but the few data are not very constraining.
$Z$ $p_T$ distribution data and top quark cross sections used are included at full NNLO.  For the inclusive $t\overline{t}$ cross sections the top mass is determined from the fit using a input pole mass value of $m_t=172.5\pm 1$ GeV, and is effectively treated as a nuisance parameter. At NNLO a value consistent with the input is obtained from the fit, while at NLO a lower value $m_t \approx 170$ GeV is preferred, the lower mass compensating for the lower hard cross section at NLO. The dependence of the PDF fit on the top quark mass (including also differential distributions) was examined in \cite{Cridge:2023ztj} and is discussed later in Section~\ref{sec:alphas}.  
For all data, electroweak  corrections are included where possible, except when these are already  subtracted from the data supplied. 

The MSHT20 PDFs have seen a very significant extension of the parameterisation. The MSTW and older PDFs were parameterised in terms of polynomials in $x^{0.5}$, with $\sim 30$ free parameters. However, \cite{Martin:2012da} demonstrated that this was not sufficiently flexible to describe Tevatron and LHC $W$ asymmetry data. It was necessary to introduce an improved basis for the interpolating function 
\begin{equation}
\label{eq:chebmmht}
I_f(x)=\sum_i^n \alpha_{f,i}T_i(y(x))\;,
\end{equation} 
where $T_i$ is a Chebyshev polynomial of order $i$ and $y(x)=1-2\sqrt{x}$. This maintains the half--integer separation in powers of $x$, as expected on Regge theory grounds, but in a more independent representation.
In order to determine how many parameters $n$ were needed a very large order polynomial was used 
to generate PDF--level pseudo--data points for the required distributions. The fractional deviation from this true PDF, as well as the fit quality  $\chi^2$, were  investigated  as a function of the number of parameters. 
At the time $n=4$ was sufficient for the precision of data available.
The improved parameterisation resulted in some improvement in the fit quality, and a significant change in the $u_V$ PDF at lower $x$.

In the MMHT14 set, this Chebyshev parameterisation was used at the input scale $Q_0^2=1\,{\rm GeV}^2$  for the $u_V$, $d_V$, $s+\overline{s}$ and light quark sea $S$ distributions. For the gluon a term with $n=2$ Chebyshev polynomial was included, but with a second term with a different low $x$ power to provide extra flexibility at low $x$. Standard polynomial parameterisations were taken for the much less constrained $s-\overline{s}$ and $\overline{d}-\overline{u}$ distributions. However, although in MMHT14 $n=4$ was sufficient, it was clear that $n=6$ is needed for much better than $1\%$ precision, and hence this is used for most PDFs in MSHT20. Additionally, $(\bar d/\bar u)$ is parameterised, instead of $(\bar d - \bar u)$, with the only constraint that $(\bar d/\bar u)\to $ constant as $x\to 0$. 
There are no restrictions applied to the MSHT20 PDFs other than the standard sum rules. Since Chebyshev polynomials oscillate between values of $\pm 1$ it is possible for PDFs to become negative. Indeed, the input gluon does become negative at some very low $x$ and $Q^2$ values. This is acceptable since this naturally occurs when performing backwards evolution, but the negative behaviour quickly disappears as PDF evolution to higher scales takes place. As $x \to 1$ the limited freedom in the parameterisation, along with the data constraints, results in any negative PDF values occurring only at very high $x$, where the PDFs values are $10^{-4}$ or less. At small $x$ the three flavours of light quark have a  common power-like behaviour as $x \to 0$, but the normalisations are allowed to differ from each other. 
This extended MSHT20 parameterisation leads to significant improvements in the global fit compared to the use of $n=4$, with overall $\Delta \chi^2 = -73$. This comes from the fit to high-$x$ fixed-target data, 
a reduced tension between E866 DY ratio data and LHC data, 
and improvement in the description of lepton asymmetry data. There is also some improvement in the fit to jet and $Z p_T$ data due to the additional flexibility in the gluon parameterisation.
Using $n=6$ for the parameterisations except for $s - \bar s$, results in an increase to 52 parton parameters. 

As we discuss later, the value of $\alpha_S(M_Z^2)$ can be determined by the fit. However, the default fit is performed with a definite value $\alpha_S(M_Z^2)=0.118$.  This acts as a consistency test, i.e. if the preferred value is incompatible with the world average value ~\cite{ParticleDataGroup:2024cfk} this would indicate some underlying issue. In fact, the preferred values at NLO and particularly at NNLO are very consistent with the world average. The default choice of $m_c^{{\rm pole}}=1.4$~GeV 
is close to optimal for fit quality, unlike previous analyses~\cite{Harland-Lang:2015qea}, 
which preferred lower values ($m_c^{{\rm pole}}\sim 1.25$~GeV). The default choice  
$m_b^{{\rm pole}}=4.75$~GeV gives a near optimal fit, and there is limited sensitivity to variations. 

A major constraint on the PDFs comes from the final HERA total 
cross section data~\cite{H1:2015ubc}, which is updated compared to in the MMHT14 fits. The effect of this update was studied in \cite{Harland-Lang:2016yfn} and found to have a small effect on the PDFs. 
However, there remain some difficulties in fitting the lower $Q^2, x$ data. Also included are the final HERA combined $\tilde{\sigma}^{\bar c c}$ and 
$\tilde{\sigma}^{\bar b b}$ data~\cite{H1:2018flt}, where the fit is not optimal, but similar results are obtained in other PDF studies. A particular additional new data set is the D0 electron/$W$ asymmetry~\cite{D0:2013lql}. $W^{\pm}$ bosons are produced preferentially 
in the proton/antiproton direction respectively, but 
the $V\!-\!A$ structure of the lepton decay means that $e^{\pm}$
is frequently emitted opposite to the $W^{\pm}$, though in practice
leptons with particular rapidity come from a range of 
$\eta^W$ values, diluting the direct constraint on PDFs at specific $x$. 
Mapping the lepton asymmetry to $W$ asymmetry requires some
PDF-dependent modeling, but provides a more direct PDF constraint, and 
MSHT20 observe a reduced uncertainty on the $d/u$ ratio compared to use of the $e$ asymmetry, especially at
very high $x$.

The MSHT20 analysis also contains a large amount of new LHC data: extremely high precision $W,Z$ ATLAS data at 7~TeV;
high precision $W^{\pm}$ data and double differential $Z$ data at 8~TeV;
CMS 8~TeV data on the $W^\pm$ rapidity distribution;
LHCb $W,Z$ higher rapidity
distributions at $7$ and $8$~TeV; CMS $W+c \,\,{\rm jets}$ data at $7$~TeV;
ATLAS high mass Drell Yan 8~TeV data;
ATLAS data on $W^{\pm}~+$  jets at 8~TeV; $Z\,\,p_T$ distributions at 8~TeV; additional $\sigma_{t \bar t}$ data points at 8~TeV along with 
ATLAS single differential  $p_{T,t}, M_{t \bar t}, y_t, y_{t \bar t}$ distributions and CMS double differential $p_{T,t},y_t$ distributions, both at 8~TeV;
inclusive jet data from ATLAS at 7~TeV and CMS at 2.76, 7 and 8~TeV.

The fit quality is generally good, but a standard statistical approach gives a poor fit quality to some data sets, resulting from
tensions between different kinematic regions (e.g. rapidity bins) or 
between different differential distributions of the same data. 
This is frequently related to modeling-type systematic 
uncertainties, particularly for the cases of jet and $t \bar t$ data \cite{Harland-Lang:2017ytb,Bailey:2019yze}, and for some data sets a smooth 
decorrelation, similar to that advocated for 8~TeV ATLAS inclusive jet data~\cite{ATLAS:2017kux}, is used.
The Hessian approach is applied to calculate the PDF errors, with the `dynamical' tolerance criterion described in \cite{Martin:2009iq} used.
This method determines the 68\% confidence level interval for the uncertainties on PDF parameters by making use of a ``tolerance'' $\Delta \chi^2$ increase from the best fit value for each individual dataset in the global fit. Once this is exceeded for any particular data set, it provides upper and lower bounds on a dataset-by-dataset level for the PDF parameters along orthogonal eigenvector directions.
Although in the fit there are 52 free PDF parameters, in the error determination some parameter directions are found to be largely redundant, leading to significant departures from quadratic $\chi^2$ behaviour. This is corrected by fixing a subset of parameters when calculating the error eigenvectors, reducing the number to 32 pairs, i.e. 64 different eigenvectors, as asymmetry is allowed. For each eigenvector $\Delta \chi^2 = T^2$, where $T^2$ is the tolerance. For most eigenvectors $T\sim 3-4$. About half of the dominant constraints on a particular eigenvector are now provided by precision 
electroweak collider data, mainly the D0 $W$ asymmetry, $7~$TeV  and $8$~TeV ATLAS $W,Z$ and CMS $W$ data.  About eight eigenvectors associated with the $\bar d /\bar u$ ratio are primarily constrained by the E866 Drell-Yan ratio. Ten eigenvectors pertaining to high-$x$ quarks are constrained predominantly by fixed target DIS data, and
the eigenvectors related to the $s, \bar s$ sum and asymmetry are still mainly constrained by CCFR, NuTeV dimuon data. 
Clearly, without making use of assumptions and/or restictive models, a completely global fit is necessary to provide a full constraint on all PDFs. 
HERA data provides good constraints on a wide variety of PDF parameters, notably those for the gluon and light sea, and particularly at small $x$, but is very rarely the best individual constraint on an eigenvector.

%----------------------------------------------
\begin{figure}[!t]
\centering
\hspace{-0.25cm}
\includegraphics[width=0.5\textwidth,trim= 0.0cm 0.1cm 0.0cm 0.2cm,clip]{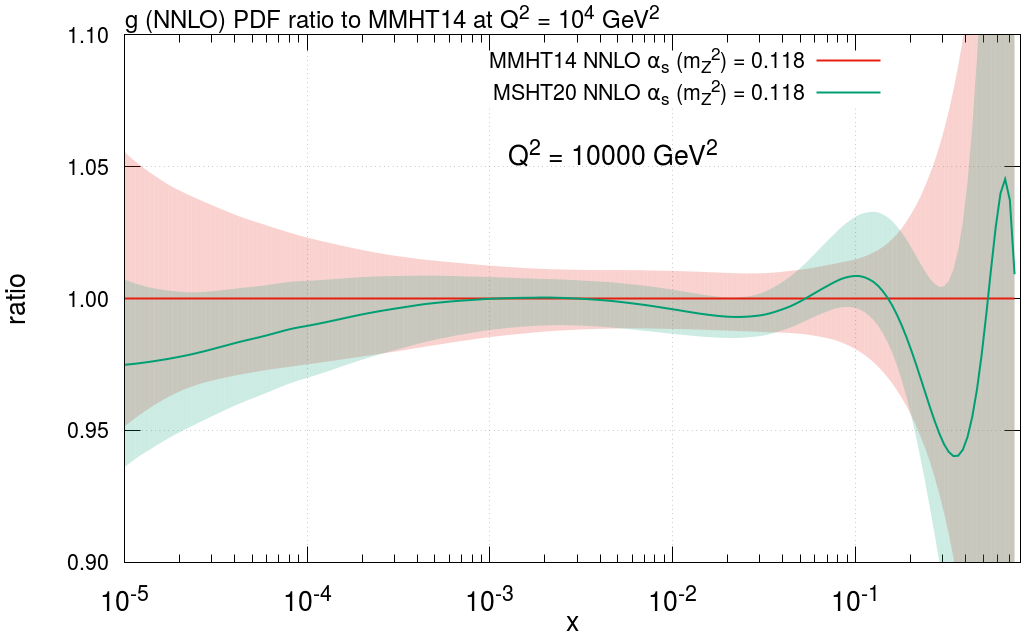} 
\hspace{-0.25cm}
\includegraphics[width=0.5\textwidth,trim= 0.0cm 0.1cm 0.0cm 0.2cm,clip]{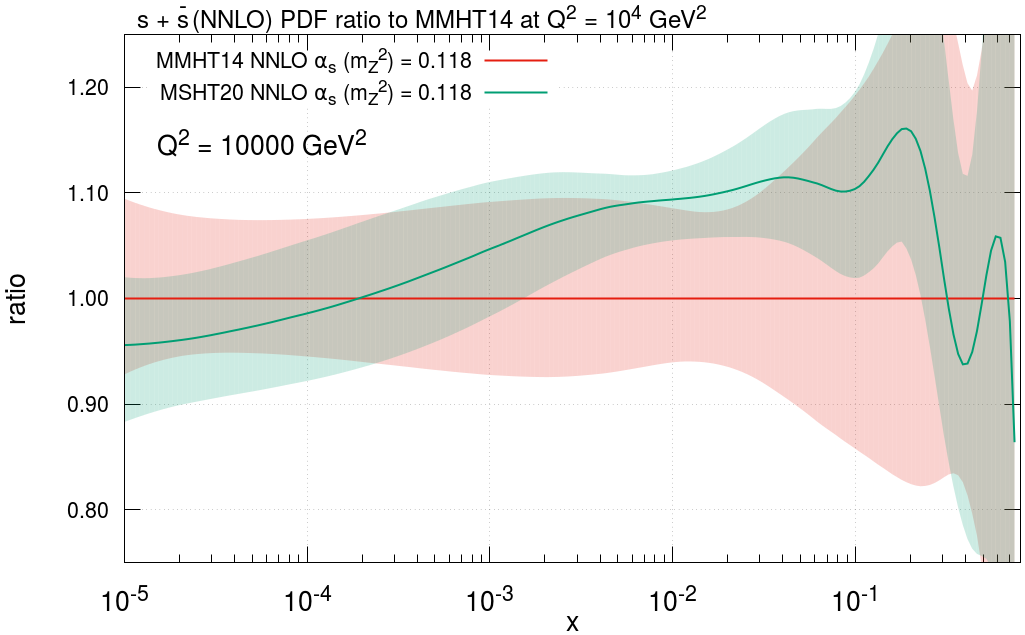}\\
\hspace{-0.25cm}
\includegraphics[width=0.5\textwidth,trim= 0.0cm 0.1cm 0.0cm 0.2cm,clip]{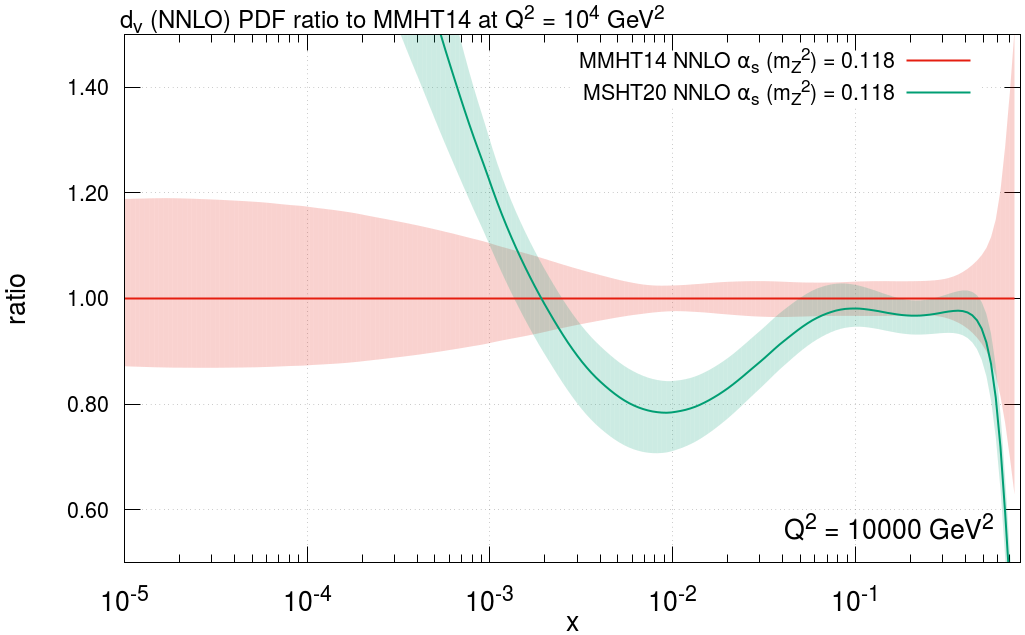} 
\hspace{-0.25cm}
\includegraphics[width=0.5\textwidth,trim= 0.0cm 0.1cm 0.0cm 0.2cm,clip]{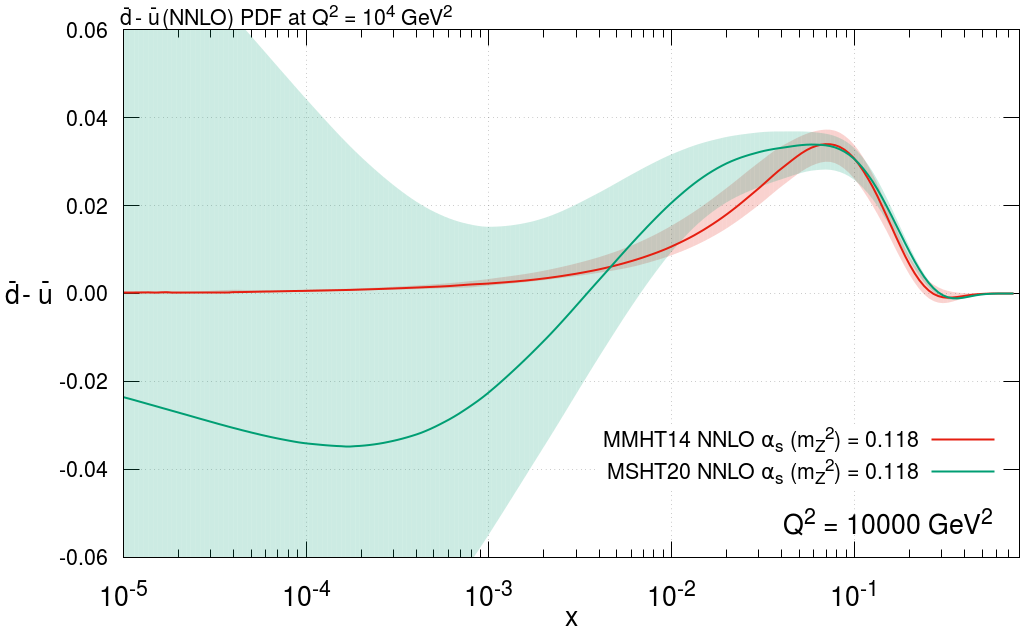}
\caption{The MSHT20 gluon (top left), strange quark sum (top right), down valence (bottom left) and $\bar d - \bar u$ (bottom right) compared to MMHT14. Figures from \cite{Bailey:2020ooq}.}
\label{fig:MSHTvsMMHT}
\end{figure}
%----------------------------------------------

The main impact of the new LHC data on the MSHT20 PDFs is from electroweak data and is on the details of flavour. We can compare the MSHT20 PDFs to those of MMHT14 in order to illustrate the changes and the degree of
stability in our PDFs.
First we show the gluon distribution in Fig.~\ref{fig:MSHTvsMMHT} (top left). The change in the central value is well within the uncertainties, but the uncertainty is reduced somewhat. The 
detail of the shape 
at high $x$ is sensitive to the LHC jet, $Z \,p_T$ and
differential $t \bar t$ data. The $Z \,p_T$ data act to pull the gluon upwards but differential 
$t \bar t$ data pulls the gluon downwards, with each affecting the lower $x$ normalisation via the momentum sum rule. There is a variation in the pull from jet data sets but the total effect is 
slightly downwards. The effect of different datasets, including those for inclusive jets, on the MSHT PDFs were examined further using the $L_2$ sensitivity method in \cite{Jing:2023isu}.
A more significant change can be observed in the strange quark which increases below $x=0.1$, 
Fig. \ref{fig:MSHTvsMMHT} (top right), due to the  
ATLAS 7, 8~TeV $W$ and $Z$ data, which all influence the PDFs in a similar fashion. There is also a very 
significant change in the shape of the valence quarks, in particular in $d_V$, caused by the LHC data on $W,Z$ but also by the extended  parameterisation freedom,
Fig. \ref{fig:MSHTvsMMHT} (bottom left). We note that valence quarks in the low $x$ region are small in absolute value and determined largely by extrapolation.
There is also some change in the weakly constrained high-$x$ region for light antiquarks, and a slight decrease in each at lower $x$  to compensate for the increase in the strange quark.
The PDF representation of the difference between the $\bar u$ and $\bar d$, shown in Fig. \ref{fig:MSHTvsMMHT} (bottom right) 
is completely modified due to the new type of parameterisation, leading to a dramatic increase in uncertainty at small $x$, and a marginal tendency for $\bar d -\bar u< 0$. However, $\bar d /\bar u$ has very small low-$x$ uncertainty and the ratio $\to 1$ as $x \to 0$ to a high level of accuracy without this being  input as a constraint.  

\section{Updates - approximate N$^3$LO and QED-corrected PDFs.}

\begin{figure}
\begin{center}
\includegraphics[scale=0.6]{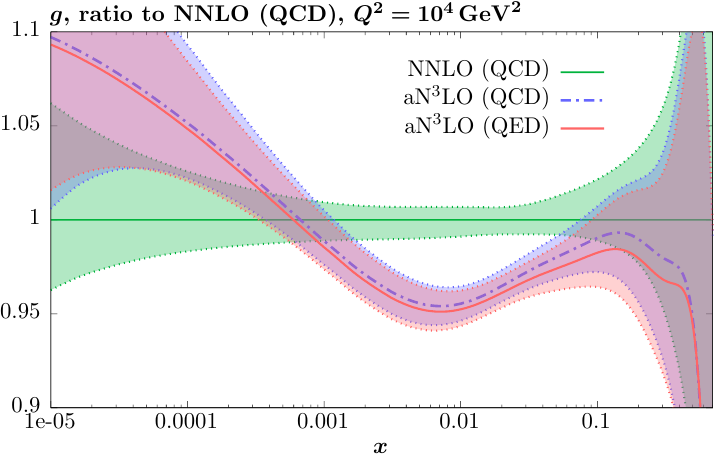}
\includegraphics[scale=0.31]{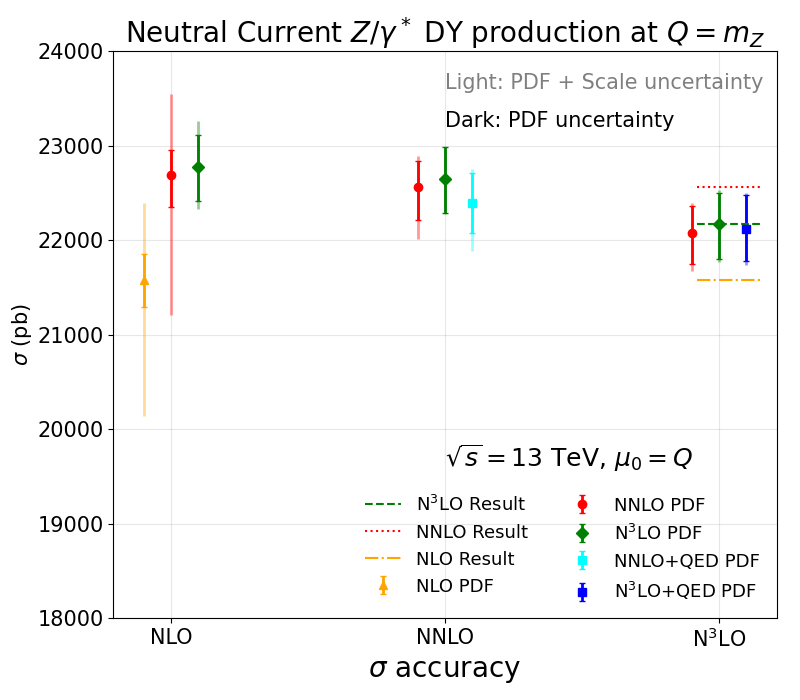}
\caption{\sf The NNLO gluon distribution compared to the aN$^3$LO QCD-only and the aN$^3$LO QCD+QED gluon distributions as a function of $x$ at $Q^2=10^4$~GeV$^2$ (left). The prediction for the cross section for production of a $Z$ boson using a variety of combinations of cross section of given perturbative order and PDFs produced at various perturbative orders (right).
The left-hand figure is from~\cite{Cridge:2023ryv}.}
\label{fig:n3logluon}
\end{center}
\end{figure}

There have been numerous updates using the MSHT20 PDFs as a baseline, e.g. dedicated studies of the sensitivity of the PDFs to different values of the heavy quark masses, and of $\alpha_S(M_Z^2)$ \cite{Cridge:2021qfd}, discussed later. 
The PDFs have also been presented with an inclusion of relevant electroweak corrections, in particular a photon parton distribution \cite{Cridge:2021pxm} using a very similar formalism to that in \cite{Harland-Lang:2019pla} (based on the formalism developed in \cite{Manohar:2016nzj,Manohar:2017eqh}). We produced the first approximate N$^3$LO (aN$^3$LO) PDFs \cite{McGowan:2022nag}, utilizing the information already available at the time on ${\cal O}(\alpha_S^4)$ corrections to splitting functions, ${\cal O}(\alpha_S^3)$ transition matrix elements and DIS coefficient functions. We parameterise missing information at N$^3$LO in terms of nuisance parameters, automatically obtaining a theoretical uncertainty. Some significant changes in the PDFs were seen at aN$^3$LO, in particular a decrease in the gluon at $x \sim 0.01$, as shown in Fig.~\ref{fig:n3logluon} (left).
A similar study has recently been performed by the NNPDF group \cite{NNPDF:2024nan}. The two evolution codes have been benchmarked, and the effects of updates \cite{Falcioni:2023luc,Falcioni:2023vqq,Moch:2023tdj} to the information available on the splitting functions examined~\cite{Cooper-Sarkar:2024crx,Andersen:2024czj}. Both groups 
see similar trends in the aN$^3$LO to NNLO comparison of PDFs, though these are generally a little smaller in the NNPDF case. In Fig.~\ref{fig:n3logluon} (right) we see the effect of the N$^3$LO correction to DY production, where the cross-section correction leads to a decrease of about 2$\%$. The inclusion of the aN$^3$LO PDFs acts slightly in the opposite direction, hence improving the perturbative stability.  
We have also made a first investigation of the effects of using dijet instead of inclusive jet data in \cite{Cridge:2023ozx}, including the case of the aN$^3$LO PDFs. At NNLO 
%we agree with the general observations of \cite{AbdulKhalek:2020jut}; 
we see less tension between dijet data and the other data in the fit than for inclusive jet data, and that the fit quality clearly improves for dijet data if N$^3$LO PDF corrections are added, which is not obviously manifest for inclusive jet case. We also investigate in more detail the impact of the $Z$ $p_T$ data \cite{Cridge:2023ozx}. 

We have repeated the inclusion of QED corrections, including a photon PDF, performed previously at NLO and NNLO~\cite{Cridge:2021pxm} (and based on the formalism developed in \cite{Manohar:2016nzj,Manohar:2017eqh}), into the aN$^3$LO study \cite{Cridge:2023ryv}. 
The QED corrections to PDFs at aN$^3$LO produce similar effects to lower orders. In particular, the photon PDF is slightly larger, and similarly to lower orders, while the QED-corrected gluon and strange quark
PDF are marginally smaller due to loss of momentum to the photon, as seen in Fig.~\ref{fig:n3logluon} (left). The high-$x$ valence quark is also a little lower due to photon radiation leading to an extremely small reduction in the prediction for Drell Yan production.
At NLO and NNLO inclusion of QED corrections leads to some deterioration in fit quality, but this becomes insignificant at aN$^3$LO. More understanding of the impact of subtly different choices for the inclusion of QED effects into the PDFs made by different groups has also been gained through comparison studies~\cite{Cooper-Sarkar:2025sqw,Cridge:2026jyh} performed on top of both NNLO and aN3LO QCD PDF fits.

\begin{figure}
\begin{center}
\hspace{-0.4cm}
\includegraphics[height=4.4cm,width=7cm,trim=0.0cm 0cm 0.4cm 0.8cm,clip]
{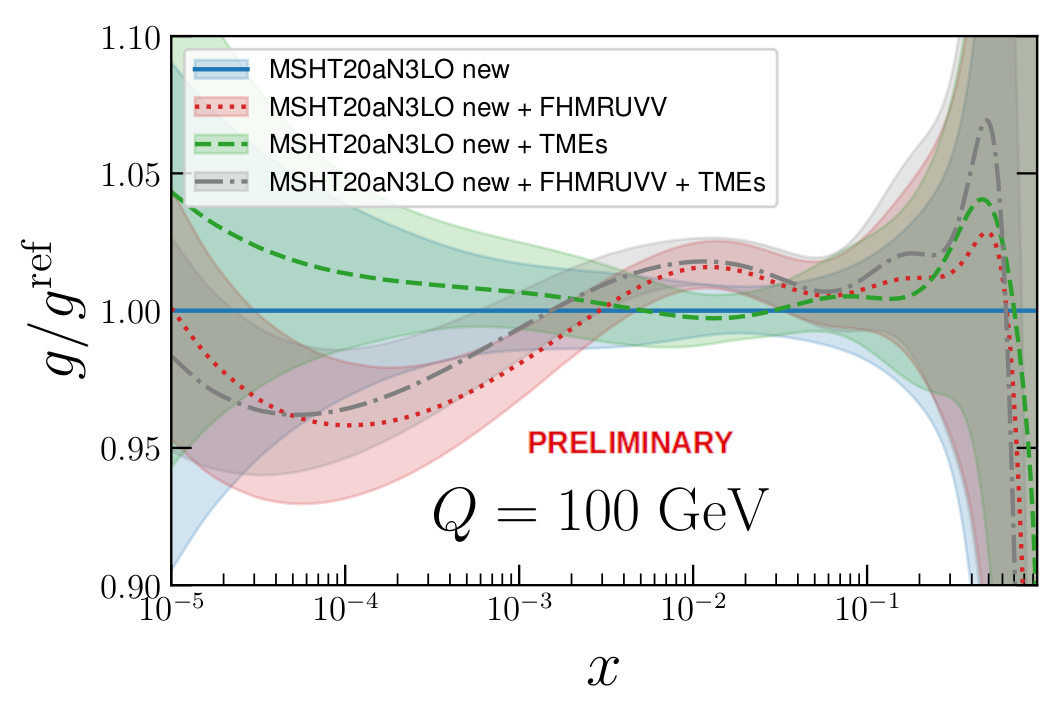}
\hspace{-0.3cm}
\includegraphics[height=4.4cm,width=7cm]
{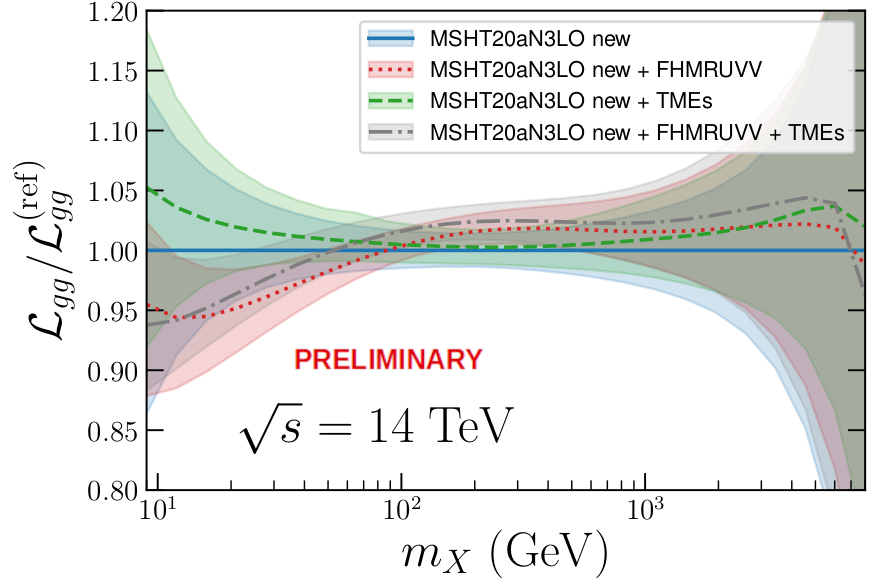}
\caption{\sf The gluon distribution at $Q^2=10^4$~GeV$^2$ (left) and gluon-gluon luminosity as a function of final state invariant mass $M_X$ (right) resulting from the MSHT global fit at aN$^3$LO with updated splitting functions and transition matrix elements compared to that from the published MSHT aN$^3$LO PDFs and to the MSHT NNLO PDFs. Figures from~\cite{DIS25proceedings}. 
}
\label{fig:newsplit}
\end{center}
\end{figure}

We have investigated the impact of including more up--to--date information on splitting functions, i.e. that contained in \cite{Falcioni:2023luc,Falcioni:2023vqq,Moch:2023tdj,Falcioni:2024xyt,Falcioni:2024qpd}, and transition matrix elements~\cite{Ablinger:2022wbb,Ablinger:2023ahe,Ablinger:2024xtt}. The effects on the gluon of a refit with the central splitting functions updated accordingly and with the updated transition matrix elements are shown in Fig.~\ref{fig:newsplit} (left), where the uncertainties 
do not currently include a component due to the remaining uncertainties on the splitting functions. The updates change the aN$^3$LO gluon a little compared to the  official
aN$^3$LO MSHT PDFs, raising it by about $1.5\%$ near $x = 0.01$. Otherwise, the features of the comparison of the aN$^3$LO PDFs to those at NNLO remain very similar. The gluon, while closer to that at NNLO than in our original 
aN$^3$LO study, still experiences a relatively large change with the increase in perturbative order, and remains inconsistent with the NNLO gluon near $x=0.01$. This may seem surprising, especially given the close similarity between the NLO and NNLO gluons. However, new dominant logarithms in $1/x$ appear at next perturbative order, i.e. for the gluon-gluon splitting function the leading term $\sim (\alpha_S^n \ln^{n-1}(1/x))/x$, so next perturbative order includes an enhancement at small $x$. Moreover, the leading terms are accidentally zero at NLO, and NNLO, but not at N$^3$LO, so 
two new dominant logarithms appear at this order (but at no subsequent higher orders), which suggests that the change in PDFs, particularly in the small-$x$ gluon, may be particularly large going from NNLO to N$^3$LO. Note that standard uncertainty estimates at NNLO, e.g. scale variation, would not anticipate the appearance of these new dominant terms at higher order.   

The $\chi^2$ for the aN$^3$LO global fit is about 125 higher than before corrections (still more than 90 units lower than at NNLO), though this would certainly improve once the remaining uncertainty in the splitting functions is allowed for~\cite{DIS25proceedings}. However, it is very likely that the 
fit quality with full N$^3$LO would be worse than our original aN$^3$LO fit as the preferred values of the nuisance parameters would likely be determined not only by the true N$^3$LO corrections but any further missing uncertainties (this would be similar for other methods of estimating uncertainty, at N$^3$LO e.g. those in  \cite{NNPDF:2019vjt}). Ideally, this should be corrected by taking into account sources of uncertainty beyond N$^3$LO, but in our initial study in \cite{McGowan:2022nag} we assumed the theoretical uncertainty was dominanted by the missing components of N$^3$LO. 
The impact of the fit with the updates on the gluon-gluon luminosity for a centre-of-mass energy of 14~TeV is shown in Fig.~\ref{fig:newsplit} (right). The change in the gluon compared to the published aN$^3$LO PDFs is diluted slightly by the integration over $x$ and the increase for $M_X=125$~GeV is a little less than 2 $\%$.  A complete update of the aN$^3$LO PDF determination is an ongoing study. 

\section{Uncertainty investigations - Closure Tests}

Given the high precision theory and wealth of data from multiple experiments we can hope to constrain PDFs with high levels of precision and accuracy. However, in order to produce a realistic uncertainty, we still have to use an inflated uncertainty and tolerance. Indeed, the PDFs from different groups only tend to agree to an acceptable level once this 
tolerance is applied. 
Each group uses similar, but not identical choices of data and theoretical ingredients, and there are more substantial differences in the methodologies, and these choices have a significant impact on the resulting PDFs and their uncertainties. We have started to address this issue directly~\cite{Harland-Lang:2024kvt}, focusing initially on the form of the PDF parameterisation. We have made the first completely like--for--like comparison between two global PDF fits, namely MSHT and NNPDF~\cite{NNPDF:2021njg}, where the difference is guaranteed to be due only to fitting methodology, and PDF parameterisation in particular.

\begin{figure}
\begin{center}
\hspace{-0.5cm}
\includegraphics[scale=0.563]{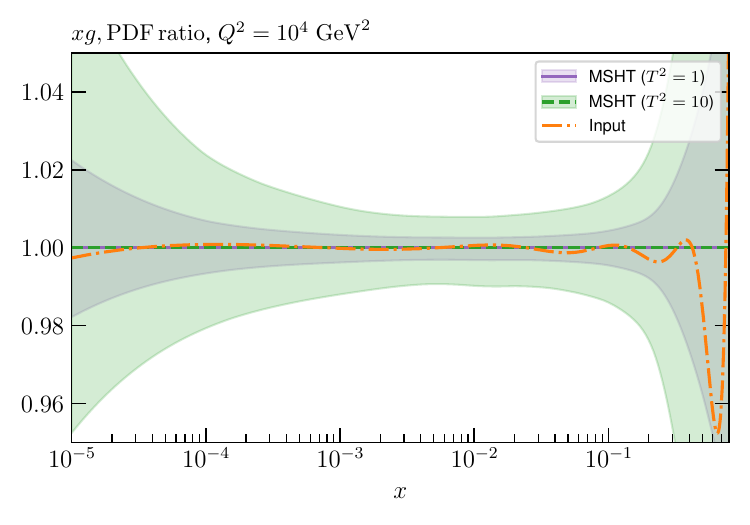}
\hspace{-0.5cm}
\includegraphics[scale=0.563]{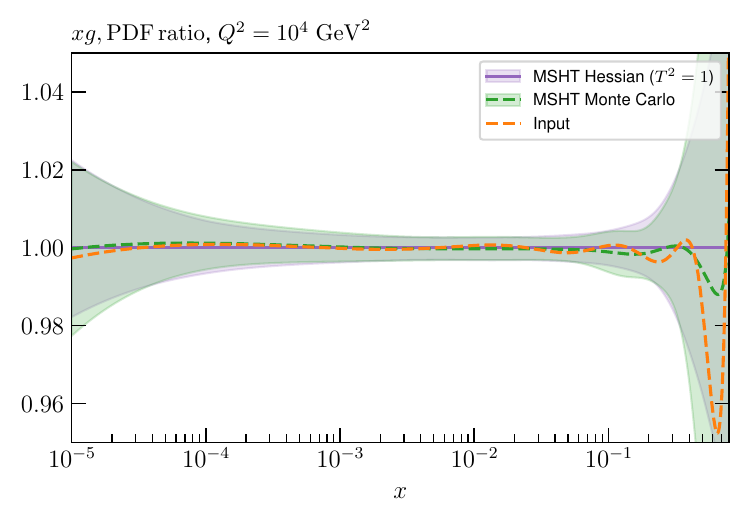}
\caption{\sf \footnotesize{The gluon PDF at $Q^2=10^4 \, {\rm GeV^2}$ resulting from an unfluctuated closure test fit to the NNPDF4.0 dataset using the MSHT20 parameterisation. The uncertainty with a $T^2=1$ ($T^2=10$) tolerance is shown in purple (green) and  the NNPDF4.0 input is the dashed orange line (left).
The same but from a fluctuated closure test fit. The Hessian PDF uncertainties with $T^2=1$ from the unfluctuated closure test are shown in purple and the result of fluctuated closure test are shown in green (right). Figures from \cite{Harland-Lang:2024kvt}.}}
\label{fig:glcl_rat}
\end{center}
\end{figure}

By developing a suitable interface to the publicly available \texttt{NNPDF} fitting code~\cite{NNPDF:2021uiq} we evaluate the fit quality and find the best fit within the NNPDF framework (using their theory and choice of dataset) but for a PDF set parameterised at input scale $Q_0$ using the basis of Chebyshev polynomials used for the MSHT20 PDFs. Two types of closure test are performed, where we generate pseudodata from the central replica of the NNPDF4.0 (perturbative charm) PDFs, and the pseudodata are unfluctuated, or are fluctuated by the experimental uncertainties. In the unfluctuated test we find a $\chi^2/{\rm N}_{\rm pts} \to 0$, showing that the PDF parameterisation is flexible enough to obtain a perfect fit. A comparison between the fit gluon PDF and the input is shown in Fig.~\ref{fig:glcl_rat}~(left) where the uncertainty using the textbook $\Delta \chi^2 = T^2=1$ and the $T^2=10$ case are indicated.
Other PDFs are similar. The agreement between the input and fit is extremely good, $\sim 10\%$  of the $T^2=1$ uncertainty in the data region. In the high $x$ region (and for poorly constrained low $x$ valence distributions) the agreement is, as expected, less good.
We also generate 100 differently fluctuated pseudodata sets and obtain a set of 100 fits. This fluctuated test assesses the PDF uncertainties that come out of the fit, i.e. we expect the input PDF set to lie within the 1$\sigma$ uncertainty band of the fit PDFs with approximately 68\% frequency. Fig.~\ref{fig:glcl_rat}~(right) shows the result from the unfluctuated fit to the 68\% C.L. ensemble of fits to the fluctuated pseudodatasets. In the data region the agreement between the MC replica and Hessian result is very good. Futhermore, the Hessian uncertainty on the 100 fluctuated fits is stable and matches the $T^2=1$ Hessian uncertainty from the unfluctuated fit. The closure test is therefore passed by each individual PDF at any given $x$ value within the data region. More quantitaive numerical comparisons, can be found in~\cite{Harland-Lang:2024kvt}.

We also fit the real NNPDF4.0 dataset, producing `MSHT fits' where both the data in the fit and the theory settings are identical to the NNPDF4.0 fit and the only difference is the input PDF parameterisation. The global fit qualities are calculated for perturbative and for fitted charm, where the charm quark PDF is parameterised using the same basis as for other PDFs in the MSHT analysis. We perform fits both with and without positivity imposed on the input MSHT PDFs. For perturbative charm we obtain $\chi^2/N_{pts}=1.240, 1.262$ without or with positivity constraint respectively compared to $\chi^2/N_{pts}=1.282$ for NNPDF4.0. For fitted charm we obtain $\chi^2/N_{pts}=1.222, 1.224$ compared to $\chi^2/N_{pts}=1.233$ for NNPDF4.0. Hence, despite the nominally greater flexibility in the NNPDF4.0 fit, the MSHT parameterisation produces a noticeably better fit quality. In the case of perturbative charm this is partially due to the PDF positivity imposed by NNPDF, mainly  of the gluon at low $x$, which is not the default in the MSHT fit, though the fit remains a little better even with this constraint. In~\cite{Harland-Lang:2024kvt} the reasons for this are considered and the possibility that the MSHT parameterisation leads to overfitting examined. No evidence is found for this and if a more restrictive parameterisation is taken the closure tests are no longer passed, and a much worse fit quality is obtained.

\begin{figure}
\begin{center}
\hspace{-0.55cm}
\includegraphics[scale=0.565]{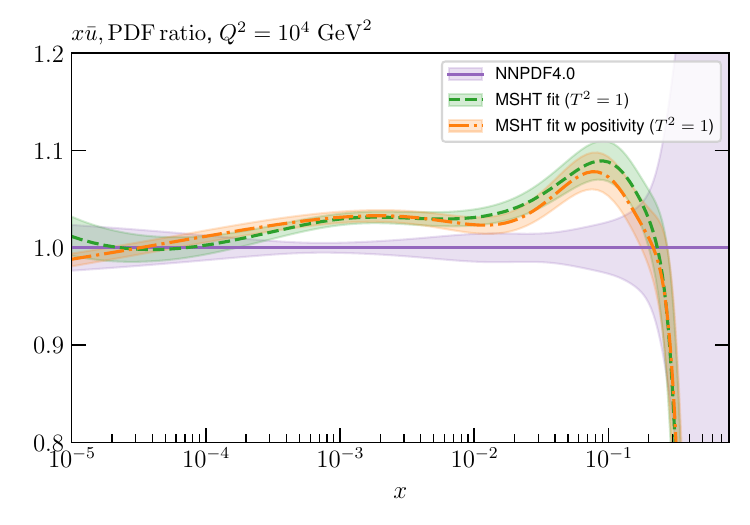}
\hspace{-0.55cm}
\includegraphics[scale=0.565]{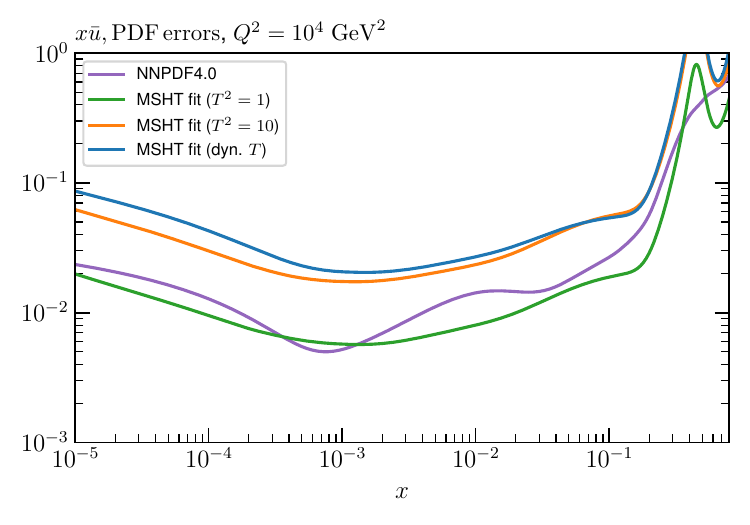}
\caption{\sf \footnotesize{ The $\overline{u}$ at $Q^2=10^4 \, {\rm GeV^2}$ from a PDF fit to the NNPDF4.0 dataset/theory (fitted charm) setting using the MSHT20 parameterisation. The ratio to the NNPDF4.0 set is shown (left), with MSHT PDF uncertainties corresponding to $T^2=1$. The corresponding PDF uncertainties are shown (right), for fixed and dynamic tolerances in the MSHT case. Figures from \cite{Harland-Lang:2024kvt}.}}
\label{fig:pdfcomp}
\end{center}
\end{figure}

We find noticeable differences in the MSHT fit PDFs, such that the central values of these are not always compatible with the NNPDF4.0 PDF within the nominal NNPDF uncertainties. These are larger in the perturbative charm fits, but are still present with fitted charm. An example is shown in Fig.~\ref{fig:pdfcomp} (left) where the ratio to the NNPDF result is shown, with $T^2=1$ uncertainties taken for the MSHT results. We also show (right) the PDF uncertainties, and the NNPDF uncertainty is similar to the MSHT case with $T^2=1$, and much smaller than with $T^2=10$. 
In this study we also investigate reasons for including a tolerance in global PDF fits, performing for the first time closure tests where data/theory inconsistencies are built in. These do not lead to any increase in the $T^2=1$ PDF uncertainty, but can lead to different subsets of data producing PDFs which are inconsistent with each other when using their nominal uncertainties.

Hence, within self-consistent closure tests the default MSHT20 parameterisation reproduces the features of the input set very well. In this context, which is idealized regarding dataset consistency and also data - theory consistency, the parameterisation is not a major source of limited precision or uncertainty for our PDFs. However, we see that different methodologies, choices of real data or data inconsistencies can lead to changes greater than the uncertainties obtained using $T^2=1$. Including theoretical uncertainties in PDF determinations \cite{McGowan:2022nag,NNPDF:2019vjt,NNPDF:2019ubu}
improves consistency between different PDFs and relieves some tensions between data sets, but only to a limited extent. Hence, investigations are ongoing to understand the source of the need for $T^2>1$ tolerance, and to try to quantify it to some extent.

\section{Improved treatment of data uncertainties and implications for tolerance}

\begin{center}
\begin{figure}
\hspace{-0.7cm}
\includegraphics[scale=0.266]{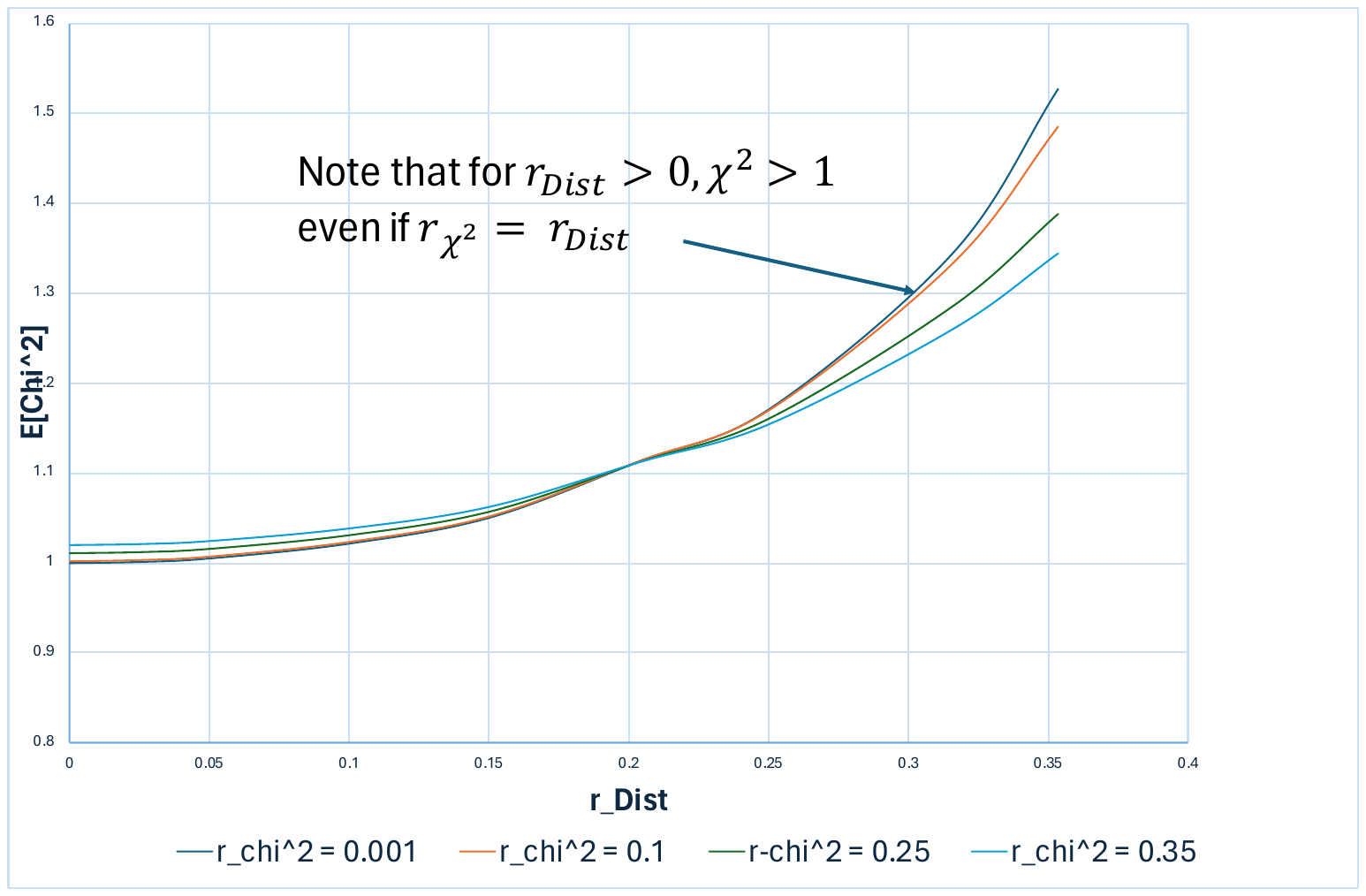} 
\hspace{-1.1cm}
\includegraphics[scale=0.266]{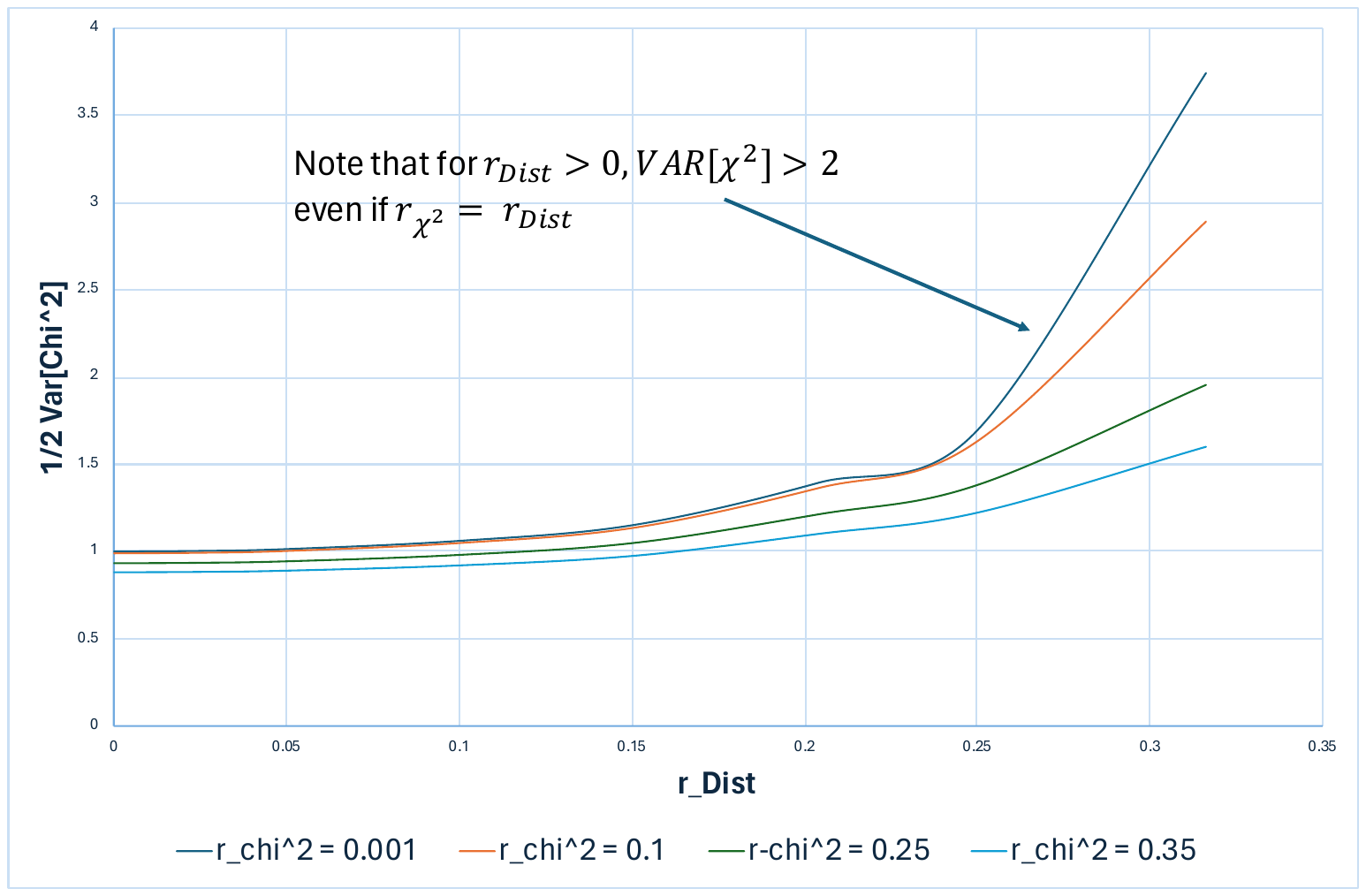}  
\caption{$E[\chi^2]$ as a Function of $r_{Dist}$ for 4 different $r_{\chi^2}$ ($\sigma_{i}=\sigma_{u_i} = 1 $ (left).  $Var[\chi^2]/2$ as a Function of $r_{Dist}$ for 4 different $r_{\chi^2}$ ($\sigma_{i}=\sigma_{u_i} = 1 $) (right). Figures from \cite{Reader:2024cbb}.}
\label{Figure 1}
\end{figure}
\end{center}

The excellent precision in PDFs when using $T^2=1$ is because experimental errors, particularly on electroweak data, are now incredibly small. The precision is now frequently limited by systematic uncertainties and the errors obtained in the PDFs extracted from these data will also be increasingly determined by experimental systematic errors. However, there are often significant uncertainties in both the size and precise degree of correlations of these systematic errors. Hence, we have been investigating an approach where "errors on errors" can be incorporated into a $\chi^2$ calculation~\cite{Cowan:2018lhq}, and studying what such a model tells us about the resulting PDF errors \cite{Reader:2024cbb}. This is based on the idea that if we have a source of systematic uncertainty $\sigma_{u_i}$, the estimated variances, $v_i$, of $\sigma^2_{u_i}$, may be modeled as Gamma distributed, and $r_i$ is then defined as the relative uncertainty in the estimate of the systematic error. This model can be reinterpreted as a Student's t-distribution. As a simple example we consider the case can where we draw our data from a distribution where statistical errors are normally distributed and uncorrelated systematic errors are t-distributed. We can investigate the expectation of fit quality $E[\chi^2]$, and the Variance of $\chi^2$, $Var[\chi^2]$. In Figure \ref{Figure 1} we plot both the $E[\chi^2]$ and $Var[\chi^2]$ as functions of $r_{Dist}$, where each has been calculated at various different $r_\chi^2$, i.e. where different assumptions about the value of $r_i$ are used in the data generation and $\chi^2$ calculation.  This confirms the expectation that a lack of understanding of the true uncertainties will tend to lead to a value of $\chi^2$ in fit greater than 1 per point, but only of the order $E[\chi^2] \sim 1.2$ (interesting this is true even if the uncertainty in the error is accounted for in the $\chi^2$ calculation). The variance of the $\chi^2$ also increases significantly, by factors of $\sim 1.5$ if the correct $\chi^2$ definition is used, but more like $\sim 2-3$ is the default $\chi^2$ assuming perfectly known uncertainties are used. Similar studies have been performed using models which relax the assumed perfect knowledge of correlations of systematic uncertainties, again with inflated $\chi^2$ values and underestimations of the uncertainties implied if the fit assumed this perfect knowledge. So far this is for hypothetical simplified cases, and some individual data sets, but not yet for full models of global PDF fits. Nevertheless, the indications are that imperfect knowledge of errors may require a tolerance of $T^2\sim 2-5$ to account for the 
assumption of perfect knowledge in the fitting procedure, i.e. not as much as often applied in PDF fits, but potentially a significant component. 

When using the PDFs for precision physics it is necessary to try to understand the full uncertainty on the PDFs, in order to minimize it, but also, importantly, to take account of it correctly. Increasingly, experimental analyses wish to utilise the constraints from new precision data yet to be included in the PDFs to further reduce the PDF uncertainties, either to examine the constraints of new data on the PDFs themselves, or to reduce the PDF uncertainty on a parameter of interest, e.g. the $W$ boson mass~\cite{CMS:2024lrd,ATLAS:2024erm}. To do so the global PDFs are ``profiled'' -- the PDF uncertainties along each PDF eigenvector are further constrained by minimising a joint $\chi^2$ of the new data and the PDFs \cite{Paukkunen:2014zia}. It is important that this profiling is done in a manner which correctly treats the uncertainties of the input PDFs and that the inherent assumptions of this approach are understood so as not to result in inconsistent or overly constrained/optimistic uncertainties. In the case of MSHT, the use of a tolerance in our PDF uncertainties means that our PDF errorbands represent not $\Delta\chi^2=1$ but rather $T^2$. The tolerance must be included in the profiling \cite{Schmidt:2018hvu, AbdulKhalek:2019mps} else the impact of the new data is over-weighted relative to the data used in the PDF extraction, resulting in an inconsistent change in PDF central value, a large increase in $\chi^2$ for existing datasets in the global fit, and too large a reduction in uncertainties. 
This would be revealed when such data are included into the PDF fits and much milder effects observed. If incorrectly profiled uncertainties are propagated to parameters of interest, the PDF uncertainty on these parameters will similarly be underestimated, and this may have a large effect on the total uncertainty in cases where it dominates. It is also important to understand that the assumptions of profiling are invalidated when the profiled PDFs are pulled too far from the global fit input or overly constrained by the new data. In such cases again false conclusions may be reached and a global PDF fit incorporating the new data is the only way to fully capture its constraints.

\vspace{-1.3cm}

\section{Determination of $\alpha_S(M_Z^2)$ in the MSHT20 fit}\label{sec:alphas}

Using the improved MSHT20~\cite{Bailey:2020ooq}
framework we undertook a follow-up study~\cite{Cridge:2021qfd} which examined the effects of varying both the strong coupling or the heavy quark masses within the global fit, updating the previous MMHT14 study~\cite{Harland-Lang:2015nxa}. In this new analysis we determined both the preferred value of $\alpha_S(m_Z^2)$ and the associated uncertainty.

\begin{figure}[htpb!]
\centering
\hspace{-0.25cm}
\includegraphics[width=0.5\textwidth]{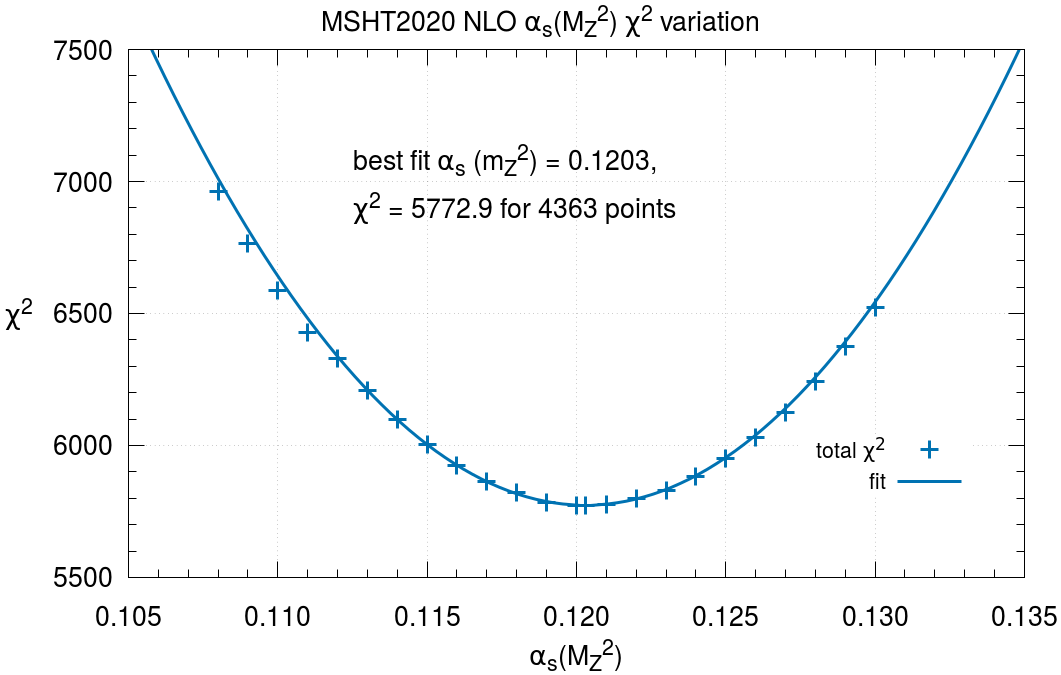}
\hspace{-0.25cm}
\includegraphics[width=0.5\textwidth]{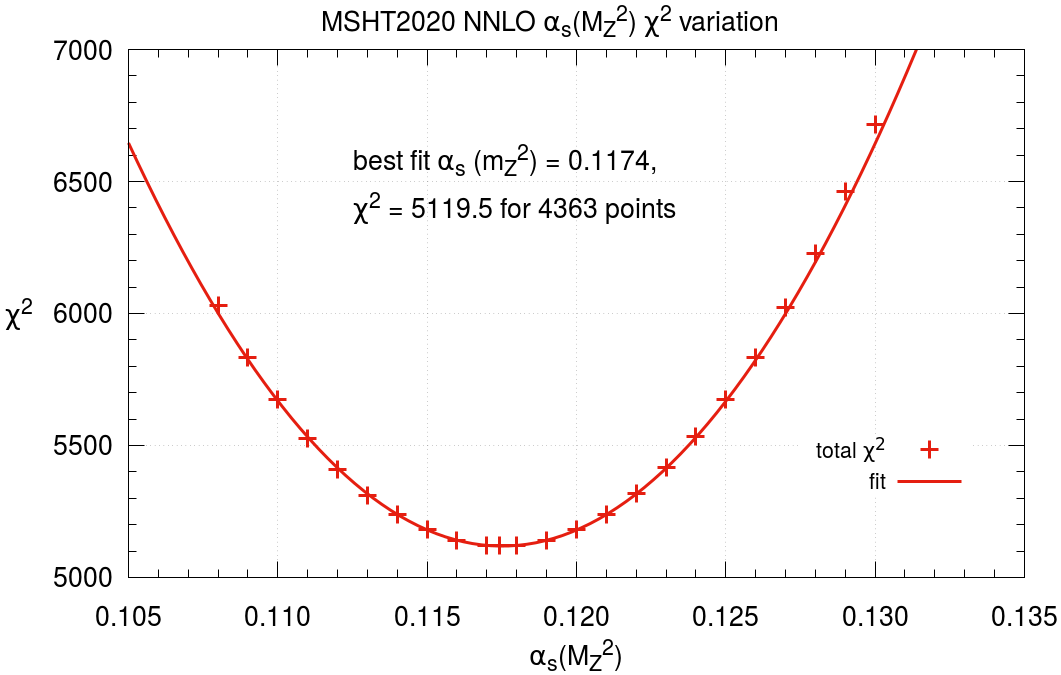}
\caption{The global $\chi^2$ as a function of $\alpha_S(m_Z^2)$ for the NLO (left) and NNLO (right) MSHT20 fits respectively. Figures originally from \cite{Cridge:2021qfd}.
}
\label{fig:alphas_msht_global}
\end{figure}

The default MSHT PDFs are provided at $\alpha_S(m_Z^2) = 0.1180$ at NNLO in order to provide a value which is consistent with the world average value of $\alpha_S(m_Z^2) = 0.1180 \pm 0.0009$~\cite{ParticleDataGroup:2024cfk} and is also a common value between different PDF fitting groups. However it is also possible to allow $\alpha_S(m_Z^2)$ to be a free parameter in the fit, with its value  determined alongside the PDFs, utilizing the global nature of the PDF fits. Using the MSHT20 framework at NLO and NNLO results in best fit values of 0.1203 and 0.1174 respectively, and provides the quadratic $\chi^2$ profiles shown in Fig.~\ref{fig:alphas_msht_global}. Hence, there is clear sensitivity to $\alpha_S(m_Z^2)$ at the global fit level.

%-----------------------------------------------------------
%-----------------------------------------------------------

\begin{figure}[htb!]
\centering
\includegraphics[width=0.87\textwidth]{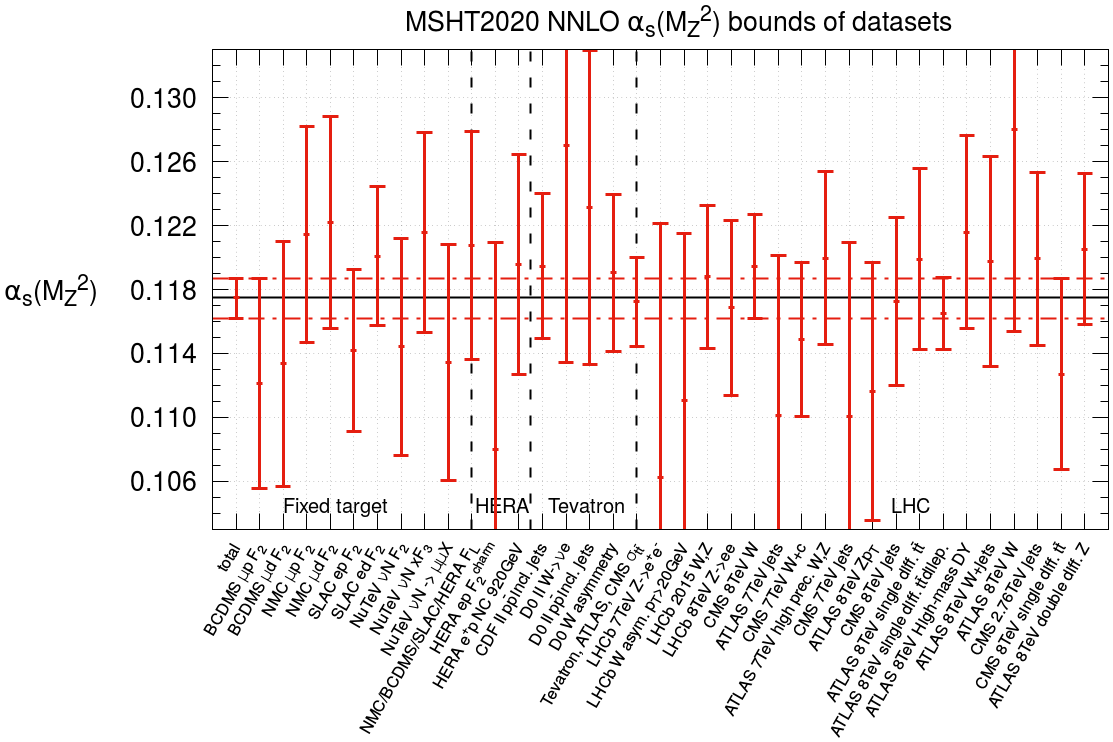}
\caption{The value of $\alpha_S(m_Z^2)$ corresponding to the best fit is shown, together
with the upper and lower $1\sigma$ constraints on $\alpha_S(m_Z^2)$, from the most constraining data sets in the NNLO analysis. The overall upper and lower bounds for $\alpha_S(M_Z^2)$ are provided by the horizontal dashed red lines. Figure originally from \cite{Cridge:2021qfd}.
}
\label{fig:alphas_msht_bounds_nnlo}
\end{figure}

In order to also use the PDF fit to determine the uncertainty we make use of the dynamical tolerance method~\cite{Bailey:2020ooq,Harland-Lang:2014zoa,Martin:2009iq} which is used for PDF parameter uncertainties. Once the 68\% confidence level interval is exceeded for any particular data set, it provides the upper and lower bounds from each dataset for $\alpha_S(m_Z^2)$. Using this criterion across the entire array of datasets in MSHT20 we obtain a bound on $\alpha_S(m_Z^2)$ from each dataset. The most relevant of these are illustrated in Fig.~\ref{fig:alphas_msht_bounds_nnlo} for the MSHT20 NNLO fit.

It is clear that different datasets prefer often significantly different optimal values of $\alpha_S(m_Z^2)$ within the context of the global PDF fit, and also that they provide very different bounds. Nonetheless, there are several general comments which can be made. In general the fixed target DIS data, most particularly the BCDMS and SLAC proton data, tend to have a preference for lower values of the strong coupling, which corresponds to a slower decrease of the structure function with increasing $Q^2$. These constraints represent one of the cleanest means of determining $\alpha_S(m_Z^2)$, as the high $x$ structure functions are dominantly non-singlet and the evolution speed is determined by the coupling alone. In contrast, the HERA deep inelastic scattering data has relatively limited sensitivity despite contributing a large number of high precision data points to the fit. This is a reflection of its role in determining the central values of the PDFs at each $\alpha_S(M_Z^2)$, meaning it is near the minimum of the profiles. As $\alpha_S(M_Z^2)$ varies the gluon distribution, particularly at small $x$, will also vary in a correlated manner to 
provide the best fit quality. The LHC data on inclusive jets and Z $p_T$ have direct sensitivity to $\alpha_S(m_Z^2)$ and these generally prefer lower values. Differential top data are omitted from the determination of the global $\alpha_S(m_Z^2)$ bounds because a fixed top mass was used in the theoretical prediction grids, and this restricts their utility as there would be correlation between $\alpha_S(M_Z^2)$ and $m_t$. A separate first study with the top mass dependence included for the top production differential data in the lepton+jet channel at 8~TeV examined this dependence on the top mass~\cite{Cridge:2023ztj}. It found that accounting for it and fitting $m_t$, $\alpha_S$ and the PDFs together can enable these datasets to produce competitive further constraints on the strong coupling as well as to simultaneously extract the top quark pole mass with a precision similar to the world average~\cite{ParticleDataGroup:2024cfk} (despite including just a subset of the top quark data). High precision W, Z data from the LHC in principle only have indirect sensitivity to the strong coupling via its impact on PDF evolution, but are still able to provide bounds on $\alpha_S(m_Z^2)$ due to their very high precision. They generally favour higher values of $\alpha_S(m_Z^2)$, albeit this is not universal.

Examining the bounds on $\alpha_S(m_Z^2)$ from each data set across the whole fit, we can extract the overall global upper and lower bounds on the strong coupling. At NNLO the BCDMS proton data provides the most constraining upper bound. This reflects its strong preference for lower values its $\Delta \chi^2$ rises rapidly as $\alpha_S(M_Z^2)$ is increased. The CMS 8~TeV W data provides the dominant lower bound, which fits the general trend of W, Z data preferring larger $\alpha_S(m_Z^2)$ values. However, the overall uncertainty is not dependent on these two data sets. Several other datasets would actually provide only slightly weaker bounds, including the SLAC proton data, for the upwards direction, and the ATLAS 8~TeV Z and high mass Drell--Yan data for the downwards direction. We obtain a symmetrized uncertainty on the strong coupling of 0.0013, which corresponds to $\Delta \chi^2 _{\rm global} = 17$, a similar change in fit quality to that determined when changing PDFs by their uncertainty along an eigenvector direction. An analogous analysis can be performed at NLO. Again, the BCDMS proton data provides the tightest upper bound (followed by the ATLAS 8~TeV Z $p_T$ and ATLAS 7~TeV inclusive jets data), whilst the lower bound is provided by the LHCb 7 and 8~TeV W,Z data (followed closely by the SLAC deuterium data). The final best fit values and uncertainties on $\alpha_S(m_Z^2)$ from MSHT fits at NLO and NNLO respectively are\footnote{The uncertainties given here are entirely experimental in nature, propagated through the PDF fit using the data point uncertainties. Theoretical uncertainties are not included within the global PDF fit in this analysis.}:
\begin{equation*}
\alpha_S(M_Z^2)_{\rm NNLO} = 0.1174 \pm 0.0013, ~~~~~~~~~~~~~ \alpha_S(M_Z^2)_{\rm NLO} = 0.1203 \pm 0.0015\,.
\end{equation*}
The NNLO result is very consistent with the PDG world average

A major benefit of performing such analyses of $\alpha_S(M_Z^2)$ within the context of a global PDF fit is the automatic inclusion of the correlations between the PDFs and the strong coupling. This is investigated in detail in~\cite{Cridge:2021qfd}, and the established method~\cite{Lai:2010nw} of treating the $\alpha_S(m_Z^2)$ uncertainty as if it is an additional eigenvector in the fit,  and consequently combining it in quadrature with the PDF uncertainty in order to obtain the total PDF + $\alpha_S(m_Z^2)$ uncertainty is illustrated. Due to these correlations, there are not only direct contributions to the total uncertainty on any observable from the strong coupling change itself, but also contributions from the corresponding correlated changes in the PDFs. The latter can sometimes counteract the direct $\alpha_S(m_Z^2)$ contribution to the uncertainty, for example in cases which are sensitive to the gluon below $x \lesssim 0.1$, where the PDF is anticorrelated with the strong coupling. In \cite{Cridge:2021qfd} studies of variations of the heavy quark mass and its correlations with $\alpha_S(m_Z^2)$ were also performed for $m_{\rm charm}$. This found only a limited correlation, the extracted value of $\alpha_S(m_Z^2)$ rising slightly with increase in the charm quark mass.

The various MSHT PDF sets are provided for public use on both the MSHT website and the LHAPDF repository 
\cite{Buckley:2014ana}. In particular, sets are provided at NLO and NNLO at fixed values of $\alpha_S(m_Z^2)$ in steps of 0.001 over the range $0.108 \leq \alpha_S(m_Z^2) \leq 0.130$. The $\Delta \chi^2_{\rm global}$ in the global PDF fit for each of these sets are provided~\cite{Cridge:2021qfd} at each fixed different value of $\alpha_S(m_Z^2)$. 

There are several developments which would help to provide improvements in the $\alpha_S(m_Z^2)$ extractions from the MSHT global PDF fits. There is clearly the question of how to incorporate theoretical uncertainties into PDF fits and the $\alpha_S(m_Z^2)$ extractions. This is being actively investigated by some of the PDF fitting groups. In particular at aN$^3$LO~\cite{McGowan:2022nag} we automatically provide a means of determining theoretical uncertainties using nuisance parameters. 
We have investigated the fit dependence on $\alpha_S(M_Z^2)$ at aN$^3$LO~\cite{Cridge:2024exf} and the higher order corrections and inclusion of 
perturbative theoretical uncertainty results in a central value which is slightly lower but consistent with NNLO and automatically leads to an increase in the overall uncertainty. The NNPDF group have been investigating uncertainty determinations via scale variations \cite{NNPDF:2019vjt,NNPDF:2019ubu}. However, current uncertainties are still limited by the available experimental data, as well as, to some extent, our understanding of its treatment. The recent LHC measurements of gluon-sensitive processes, e.g. inclusive jets and/or dijets, and top processes, have enabled improvements of our constraints on both the PDFs and $\alpha_S(m_Z^2)$. The availability of precise grids for the theoretical predictions for $\alpha_S(m_Z^2)$ sensitive processes, e.g. triple differential jet data, will enable further progress here. This could be possibly be complemented by the inclusion in the global fits of DIS jet data from HERA. Correctly accounting for the top quark mass dependence in the full variety of differential data top quark in the global fits, using the theoretical grids provided at different fixed values would provide further improvements~\cite{Cridge:2023ztj}. It would be particularly interesting to examine ratios of measurements to provide more direct sensitivity to $\alpha_S(m_Z^2)$, in particular the ratio of the inclusive 3- and 2-jet cross-sections~\cite{CMS:2013vbb}. 
The two quantities are highly correlated (both experimentally and theoretically), but this would likely cancel to some extent in the ratio, so this measurement could potentially improve the $\alpha_S(m_Z^2)$ sensitivity. However, this is not completely straightforward and requires both the experimental data and the theoretical predictions to be provided in a form suitable for inclusion in the PDFs. 

As mentioned earlier, as data and theoretical precision have increased, and statistical uncertainties have reduced, the systematic uncertainties and their correlations have become far more important in precision measurements. Global PDF fits have increasingly run into issues when using these correlated systematic uncertainties,  and an improved understanding of the precise nature and limitations of correlations would undoubtedly also benefit $\alpha_S(m_Z^2)$ extractions performed together with PDF extractions. However, more generally, we have emphasized the wide array of different datasets with sensitivity to the strong coupling used in the global PDF fits. Older fixed target data can provide the tightest bounds in some cases while newer LHC inclusive jet and high precision Drell--Yan data can  provide the dominant bounds in others. Hence, it is clear that the continual expansion of the number and perhaps more importantly, breadth of datasets included is fundamentally important for further reducing both PDF and $\alpha_S(M_Z^2)$ uncertainties. 

\section{Summary}

The MSHT PDFs are the latest in a long series of PDFs obtained from fully global fits to data which involve hadrons as at least one of the colliding particles. We have outlined the improvements inherent in the MSHT study compared to the previous iterations, in particular the last full update, MMHT2014. Many PDFs have remained very stable over time, but improved flexibility and new data lead to changes in flavour decomposition, most particularly in the small-$x$ valence quarks.  We have then focused on specific investigations based on the underlying MSHT20 baseline. We summarise the first aN$^3$LO PDF fits, and highlight and discuss the major changes compared to NNLO, in particular the gluon distribution, as well as outlining the additional inclusion of QED effects. We then summarise two recent studies attempting to understand better the source of uncertainties in PDFs.  Closure tests, i.e. producing PDFs using exactly the same data and theoretical calculations as NNPDF, demonstrate that we are able to obtain 
extremely similar PDFs to NNPDF if we use pseudodata generated by their PDFs
and the standard $\Delta \chi^2=1$ prescription for uncertainties. Hence, our parameterisation is not a limitation in precision. However, fitting to real data we find noticeable differences in the best fit PDFs and their uncertainties. We have therefore undertaken further studies to investigate this. One such difference is our requirement for a tolerance $\Delta \chi^2 \sim 10$ for the determination of the PDF uncertainties. We have performed studies analysing the implications of allowing for errors on experimental errors, and in relaxing the very strong correlations often assumed in systematic uncertainties. These have suggested such factors may explain some, though not all, of the tolerance we apply (though we emphasise that these issues will be present in all PDF fits). Finally, we summarise determinations of the strong coupling constant in our fits, illustrating the tensions that exist between the preferred values for different data sets, demonstrating that NNLO fits prefer a lower value $\alpha_S(M_Z^2)=0.117-118$ compared to NLO fits, but that these values have remained quite stable to fit updates. We also discuss how future data and improvements in the fit can lead to better determinations of $\alpha_S(M_Z^2)$.      

\section*{Acknowledgements}

T.C. acknowledges that this work has been supported by funding from Research Foundation-Flanders (FWO) (application number: 12E1323N) and by the Royal Society through Grant URF$\backslash$R1$\backslash$251540.  L. H.-L. and R. S. T. thank the Science and Technology Facilities Council (STFC) for support
via grant awards ST/T000856/1 and ST/X000516/1.

\end{document}